\documentclass[pdflatex, sn-mathphys-num, 12pt]{sn-jnl}

\usepackage{graphicx}%
\usepackage{amsmath,amssymb}
\usepackage{mathrsfs}%
\usepackage{xcolor}%

\usepackage{booktabs}%

\usepackage{xspace}

\usepackage[version=4]{mhchem}
\usepackage{todonotes}

\let\oldAA\AA
\renewcommand{\AA}{\text{\oldAA}}
\newcommand{\uu}[1]{\ensuremath{\,{\text{#1}}}}

\newcommand{\etal}{\emph{et al.}\xspace}

\usepackage[normalem]{ulem}

\begin{document}

\title{DFT and MLIP study of solute segregation to coherent and semi-coherent $\alpha$-\ce{Fe}/\ce{Fe3C} interfaces}

\author*[1]{\fnm{Amin} \sur{Reiners-Sakic}}\email{amin.reiners-sakic@unileoben.ac.at}

\author[1]{\fnm{Ronald} \sur{Schnitzer}}\email{ronald.schnitzer@unileoben.ac.at}

\author[1]{\fnm{David} \sur{Holec}}\email{david.holec@unileoben.ac.at}

\affil[1]{\orgdiv{Christian Doppler Laboratory for Knowledge-based Design of Advanced Steels, Department of Materials Science}, \orgname{Montanuniversität Leoben}, \orgaddress{\street{Franz Josef-Straße 18}, \city{Leoben}, \postcode{8700}, \country{Austria}}}

% \begin{frontmatter}

% \author[inst1]{Amin Reiners-Sakic}

% \affiliation[inst1]{organization={Department of Materials Science, Montanuniversität Leoben},%Department and Organization
%             addressline={Franz Josef-Straße 18}, 
%             city={8700 Leoben}, 
%             country={Austria}}
% \affiliation[inst2]{organization={Christian Doppler Laboratory for Knowledge-based Design of Advanced Steels, Department of Materials Science, Montanuniversität Leoben},
%             addressline={Franz Josef-Straße 18}, 
%             city={8700 Leoben}, 
%             country={Austria}}

% \author[inst2]{Ronald Schnitzer}
% \author[inst2]{David Holec}

\abstract{
Solute segregation to interfaces significantly impacts material behavior. A large majority of theoretical works focus on grain boundaries and coherent interfaces. Studies on semi-coherent interfaces are usually prohibited by the structural complexity, yielding models beyond the practical capability of density functional theory (DFT), or chemical complexity, restricted by the availability of (classical) interatomic potentials. 

This work investigates solute segregation to the coherent and semi-coherent $\alpha$-Fe/\ce{Fe3C} interface in pearlite and its effect on mechanical properties using novel universal machine learning interatomic potentials (uMLIPs). DFT calculated solution enthalpies, segregation energetics, and changes in cohesion at the coherent interface are used to benchmark several state-of-the-art uMLIPs. We find that the GRACE-2L-OAM and GRACE-2L-OMAT models most accurately reproduce the quantum-mechanical predictions.

While Cu has the strongest segregation energy of $\approx -0.3\,\text{eV}$ to the coherent interface among the investigated tramp and trace elements, all of them, As, Cr, Cu, Mo, Ni, P, Sb, and Sn, exhibit significantly more negative segregation values reaching below $\approx -1.5\uu{eV}$ in the presence of the misfit dislocation at the semi-coherent interface. The deepest traps are identified in the vicinity of the dislocation core, although the spatial distribution of segregation energies differs markedly among the solute species. The cohesion of the coherent interface is strongly reduced by Sb, Sn, P, and As, and only mildly by Cu, whereas Ni shows a negligible effect, and Cr and Mo slightly enhance cohesion. In contrast, all investigated solutes (except for P) tend to embrittle the semi-coherent interface, with Sn and, especially, Sb having the strongest impact in tensile tests performed in the out-of-plane direction.

This study highlights the importance of considering the structural features of real microstructures and paves the way for comprehensive materials design through atomistic simulations.}




\maketitle
% \begin{keyword}
% Pearlite \sep Coherent and semi-coherent interfaces \sep Segregation \sep Cohesion \sep Universal machine learning interatomic potentials \sep DFT
% \end{keyword}

% \end{frontmatter}

\section{Introduction}
Pearlite, formed through the eutectoid transformation of austenite, ranks among the decisive technological microstructures in steel. It is composed of the soft and ductile body-centered cubic ferrite ($\alpha\text{-}$\ce{Fe}, space group $Im\bar3m$) and the rigid orthorhombic cementite ($\theta
\text{-}\ce{Fe3C}$, space group $Pnma$) phase. The morphology of pearlite, which is highly influenced by its thermal history, is typically lamellar in a majority of industrial applications. The thickness of these lamellae plays a crucial role in adjusting the steels' strength and ductility, as described by the Hall-Petch relation~\cite{hall_deformation_1951}. Another critical factor influencing the mechanical properties of pearlitic steels is their chemical composition. Here, most of the literature focuses on the effect of alloying elements on the prior austenite grain size~\cite{lee_predictive_2013}, their impact on the eutectoid transformation point~\cite{honjo_effect_2016}, and \ce{C} diffusion~\cite{honjo_effect_2016, babu_diffusion_1995}, which influence the size of the pearlite colonies and lamella fineness~\cite{honjo_effect_2016}. For instance, \ce{Cr} decreases the lamellae spacing by elevating the equilibrium eutectoid transformation temperature, which enables greater undercooling, and also by lowering the \ce{C} diffusion coefficient in austenite~\cite{honjo_effect_2016}. According to~\cite{Capdevila2005-ve}, elements such as \ce{Cr} and \ce{Mo}, which stabilize ferrite and form carbides, generally lessen the spacing of lamellae, whereas those that stabilize austenite typically enlarge it. Alloying elements may also influence the lamellae's appearance. While \ce{Cu} has been shown to decrease the regularity, \ce{Ni} has been shown to have the opposite effect~\cite{Houpert1997-xk}. 

Despite the large interfacial area between ferrite and cementite in pearlite, comparatively little attention has been paid to the influence of alloying and trace elements on interface properties. One notable exception is \ce{H} embrittlement, where both experimental~\cite{Yu2019-cu, Niu2024-vz} and theoretical~\cite{Chen2023-li, McEniry2018-sq} studies have explored \ce{H} segregation and its effect on the ferrite–cementite interface. In contrast, solute segregation to pearlitic interfaces remains understudied, especially when compared with the extensive literature on segregation to grain boundaries (GBs) in $\alpha$-Fe~\cite{mai2022segregation, mai2023phosphorus, Sakic2024-ny, Reiners-Sakic2025-cq, kostwein2025tracking, Lejcek2017-ow}.

One possible reason is that solute segregation at ferrite–cementite interfaces is weaker than that at $\alpha\text{-}\ce{Fe}$ GBs, leading to smaller interfacial solute excess values that are challenging to quantify using even high-resolution techniques such as atom probe tomography (APT) or high-resolution transmission electron microscopy (HRTEM). For instance, density functional theory (DFT) calculations report segregation energies of \ce{H} to pearlitic interfaces in the range $0.12 \geq E_\text{seg} \geq -0.38\,\text{eV}$~\cite{Chen2023-li}, whereas values to $\alpha\text{-}\ce{Fe}$ GBs are more negative ($-0.34 \geq E_\text{seg} \geq -0.48\,\text{eV}$)~\cite{A_S_Kholtobina}, indicating stronger segregation tendencies to GBs. Furthermore, it may be difficult to accurately measure the interfacial excess between matrix and nm-sized \ce{Fe3C} precipitates in the case of tempered martensitic steels.

These experimental limitations can be overcome by atomistic modeling approaches, such as DFT or molecular statics. However, existing theoretical studies have primarily focused on undecorated interfaces~\cite{Zhang2015-jt, Guziewski2016-zp} or the effect of \ce{H}~\cite{Chen2023-li, McEniry2018-sq}. This is likely related to the high computational cost of DFT for such structurally complex and magnetic interfaces, which restricts simulations to small, coherent interfaces. Experimental studies also often face some constraints and, for example, APT investigations typically report \ce{H} distributions only at coherent pearlitic interface regions~\cite{Niu2024-vz}. Yet, observations in cold-drawn pearlitic steels reveal \ce{H} trapped at interfaces containing crystalline defects such as misfit dislocations~\cite{Yu2019-cu}, highlighting the importance of considering non-coherent and defect-rich interface regions. Molecular static/dynamics simulations can, in principle, handle larger and even non-coherent interfaces, but their reliability strongly depends on the accuracy of the interatomic potential. Unfortunately, only a limited number of classical potentials describe $\ce{Fe}\text{-}\ce{C}$ interactions, and even fewer capture ternary $\ce{Fe}\text{-}\ce{C}\text{-}\ce{X}$ interactions, where even less of them were fitted to actually treat these pearlitic interfaces. We are not aware of any existing interatomic potential that accounts for the wide range of alloying elements present in modern steels. Moreover, the ongoing transition of the steel industry from ore-based production to scrap-based recycling further increases the variety of elements, so-called tramp elements, introducing solutes such as \ce{Sn}, \ce{Sb}, \ce{As}, and \ce{Cu}~\cite{Dworak2023-fi, Daigo2021-lm}, in addition to classical trace elements like \ce{P}. These solutes are well known to segregate strongly to bcc-\ce{Fe} GBs, where they reduce the GB cohesion and promote intergranular fracture~\cite{mai2023phosphorus, Sakic2024-ny}. It is therefore of both scientific and industrial relevance to determine the extent of their segregation to $\alpha$-\ce{Fe}/\ce{Fe3C} interfaces and, if segregation occurs, to assess their effect on interfacial cohesion. A comprehensive investigation, therefore, requires combining the strengths of DFT calculations, namely their ability to treat chemically complex environments with quantum-mechanical accuracy, with molecular statics (and dynamics) simulations, which can treat much larger systems, including non-coherent interfaces. Recent advances in machine learning, in combination with extensive material databases obtained from DFT calculations, have enabled the development of universal machine learning interatomic potentials (uMLIPs). These potentials are capable of combining the best of both worlds: the chemical versatility of DFT with model sizes of molecular statics. Currently, the Matbench Discovery repository~\cite{Riebesell2025-ar} lists and compares the performance of over 31 different models. Still, the accuracy of these uMLIPs needs further testing, especially when it comes to structurally and chemically (defected) complex systems, i.e., going beyond defect-free ordered bulk systems. Comprehensive comparisons of this kind were performed for surface slab systems~\cite{Focassio2024-us}, defect energetics, phonon vibrational modes, and ion migration barriers for different elements and binary compounds~\cite{Focassio2024-us, Deng2025-ec, Shuang2025-lq}. However, these studies focus more on overall performance rather than seeking the best model for investigating specific materials science-related questions. Especially, Fe-based alloys (steels) deserve more attention due to their economic importance on the one hand, and non-trivial interplay with magnetism on the other hand.

Therefore, in this work, we employ DFT to compute the segregation propensities and impact on interface strength of common alloying elements such as Ni, Cr, Mo, as well as tramp and trace elements, As, Cu, Sb, Sn, and P to the $\alpha$-Fe/\ce{Fe3C} pearlite interface. The obtained results are compared with segregation energies, interface cohesion, and solution energies in bulk bcc-Fe and \ce{Fe3C} calculated using seven different uMLIPs. We have chosen CHGNet~\cite{Deng2023-tu} because, unlike other models, it is also pre-trained to explicitly predict magnetic moments. An additional six models are taken from the GRACE family~\cite{Bochkarev2024-wn} of uMLIPs, namely GRACE-1L-MP-r6, GRACE-2L-MP-r6, GRACE-1L-OAM, GRACE-2L-OAM, GRACE-1L-OMAT, and GRACE-2L-OMAT. These represent comparably new uMLIPs, where especially the GRACE-2L-OAM is one of the top performers on the Matbench Discovery scoreboard~\cite{Riebesell2025-ar}. The best model is then used to extend the interface model laterally, thereby creating a semi-coherent interface containing a misfit dislocation. Finally, the full segregation spectra are obtained for the semi-coherent interface (whose complete treatment is computationally prohibitive at the DFT level), showing significant differences to the coherent case. This indicates that the strain field around the dislocation core introduces deep traps for solute segregation. The present study paves the way for computational studies beyond traditional interatomic potentials and \emph{ab initio} methods, and showcases applications of readily available uMLIPs to predictions in complex, technologically relevant microstructures, whereby setting the baseline for experimental investigations.

\section{Methodology}
\label{sec:methodology}
Density functional theory (DFT)~\cite{hohenberg1964density, PhysRev} calculations were performed with the Vienna Ab-initio Simulation Package (VASP)~\cite{Kresse1996-gt, Kresse1996-tg} using the projected augmented wave (PAW) method~\cite{Kresse1999-if} for the electron-ion interactions and the generalized gradient approximation (GGA) within the Perdew, Burke, Ernzerhof (PBE) parameterization~\cite{Perdew1996-vd} for the exchange-correlation potential. The pseudopotentials used in this study (PAW-PBE, version 6.4) for Fe, Cr, Mo, and Ni included the $p$ semi-core electrons in the valence treatment (e.g., Fe\_pv), whereas for all other elements the recommended standard pseudopotentials were employed. A plane wave cutoff energy of $500\uu{eV}$ was chosen for the expansion of the wave functions. The first Brillouin zone was sampled with a $\Gamma$-centered mesh obtained from the automatic $k$-mesh generation method with the length parameter ($R_k$) set to $50\uu{\AA}$. All calculations regarding the bulk systems were performed using an energy-based convergence criterion set to $10^{-4}\uu{eV}$, with full structural relaxation of both atomic positions and cell degrees of freedom. This, together with the plane wave cut-off energy and $k$-mesh, should guarantee the total energy changes below $\approx 1\uu{meV/at.}$ and ensure that all force components converge to $\lessapprox 0.01\uu{eV/\AA}$. The relaxed lattice parameters of $\alpha$-Fe and \ce{Fe3C} are provided in Supplementary Materials Table~S2. Slab calculations, including surface and interface slabs, were performed using a stricter force-based convergence criterion for the ionic relaxation set to $0.01\uu{eV/\AA}$ while fixing the cell volume and cell shape. All DFT calculations were performed in the collinear spin-polarized mode. More information on the parameters used to relax the pristine interface can be found in Supplementary Materials Table~S4. 

The atomistic simulations using various uMLIPs were implemented in the pyiron simulation environment~\cite{Janssen2019-vo}.
The stopping criterion for the molecular statics calculations utilizing the uMLIPs was set to a maximum interatomic force of $10^{-4}\uu{eV/\AA}$ applied during structural optimizations where all degrees of freedom, including atomic positions, cell shape, and lattice parameters along periodic-boundaries, were optimized. For general information on the used uMLIPs, we refer the reader to the Matbench Discovery scoreboard\footnote{\url{https://matbench-discovery.materialsproject.org/}} and the GRACE website\footnote{\url{https://gracemaker.readthedocs.io/en/latest/gracemaker/foundation/}}.
During the simulated tensile tests, as counterparts of static works of separation studies (see Sec.~\ref{sec:semi-coherent}), the surface atoms were fixed and used to prescribe the elongation. Starting from the fully relaxed model, the distance between top most cementite and the bottom-most ferrite plane was elongated by $0.1\uu{\AA}$, after which all but the surface atoms were relaxed. Subsequently, the separation between bottom and top planes was again enlarged by $0.1\uu{\AA}$, and atoms in-between relaxed. This process was repeated until the overall elongation reached $10\uu{\AA}$ ($\varepsilon\approx0.25$ as the initial distance between the two surfaces in our model was $\approx40\uu{\AA}$). The resulting energy--strain curves were evaluated for the maximum energy increase (per area) before cleavage, labeled as modulus of toughness, $U_\text{t}$.

\subsection{Structural models}
\begin{figure}[ht!]
    \centering
    \includegraphics[width=0.6\textheight]{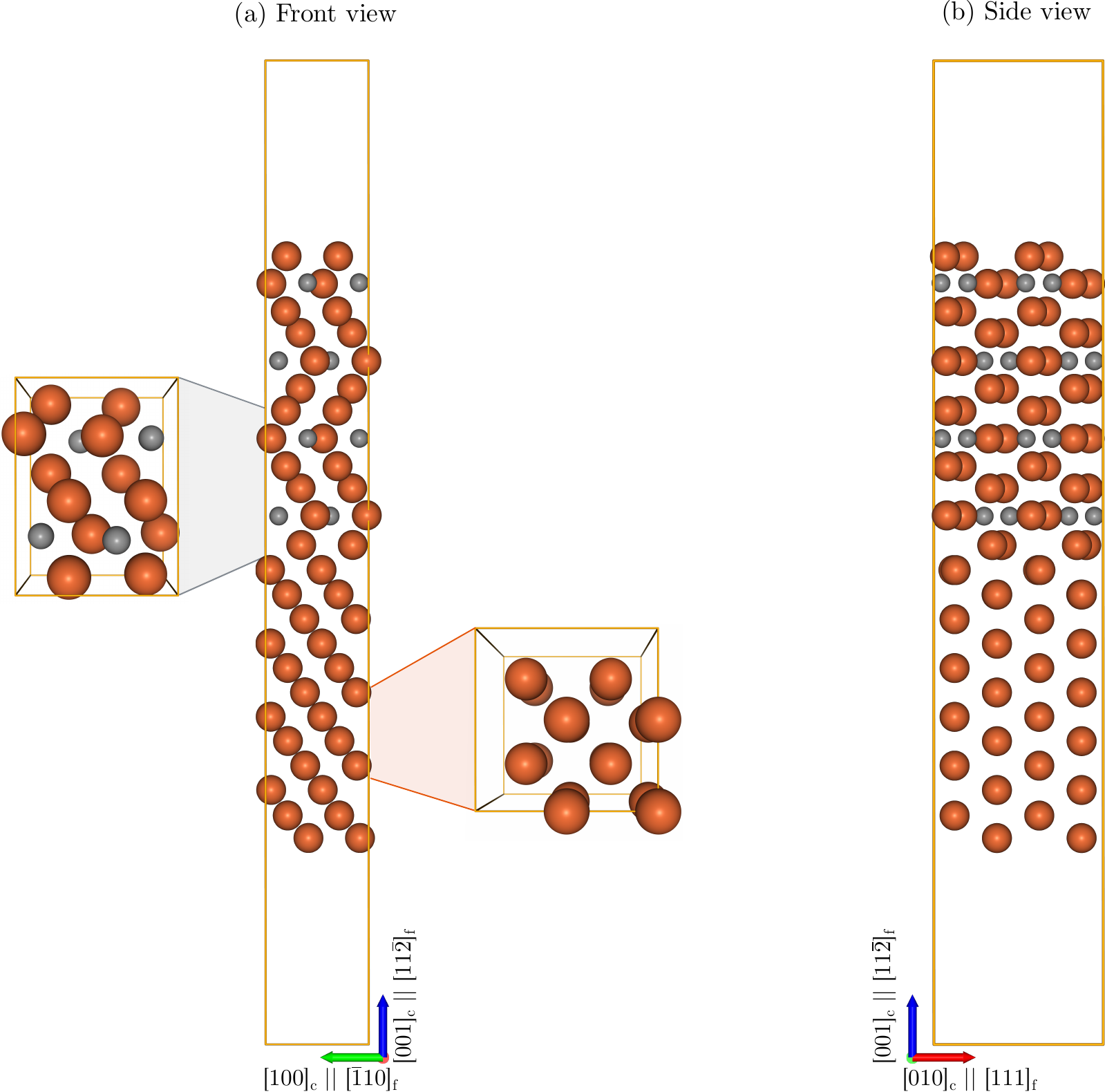}
    \caption{Coherent $\alpha$-Fe/\ce{Fe3C} interface structure with the Bagaryatsky OR. (a) Front view of the interface, including the bulk \ce{Fe3C} unit cell (left) and the bulk $\alpha$-Fe unit cell (right). (b) Side view of the interface.}
    \label{fig:atomistic_models}
\end{figure}

Although the number and relevance of different orientation relationships (ORs) between ferrite and cementite have varied over time, the Bagaryatsky OR has been reported most consistently, both in pearlitic~\cite{Zhou1992-jl} and in non-pearlitic steels containing cementite precipitates, e.g., tempered martensite~\cite{Saha2016-nm}. Therefore, in this case study we focused exclusively on the $\alpha$-\ce{Fe}/\ce{Fe3C} interface constructed according to the Bagaryatsky OR~\cite{Zhou1992-jl}, expressed as
\begin{equation*}
\begin{aligned}
\left[100\right]_{\text{c}} &\parallel \left[\bar{1}10\right]_{\text{f}}\\
\left[010\right]_{\text{c}} &\parallel \left[111\right]_{\text{f}}\\
\left(001\right)_{\text{c}} &\parallel \left(11\bar{2}\right)_{\text{f}}
\end{aligned}
\qquad\text{(Bagaryatsky OR)}
\end{equation*}
where the last line indicates the habit plane of the interface and the subscript `c' (`f') refers to cementite (ferrite).
For the construction of the interface models, the in-plane lattice parameters of cementite were strained to match those of bulk ferrite, obtained from a full structural relaxation. This choice reflects the typical morphology of pearlite, where ferrite lamellae are generally several times thicker than those of cementite~\cite{aranda2014effect}. Consequently, although cementite is elastically stiffer than ferrite (as indicated by the comparison of their elastic constants~\cite{Souissi2015-mn, Sakic2022-tt}), the lattice mismatch is expected to be accommodated primarily within the cementite. The ferrite atoms at the interface are constrained by the much thicker ferrite lamella. This effect becomes even more pronounced in non-pearlitic steels, where cementite precipitates are embedded in a ferrite matrix. The number of ferrite and cementite layers in the slabs containing an interface was set based on surface energy convergence tests. All slab calculations were performed with a vacuum layer of $20\uu{\AA}$ separating the slab from its periodic image. This setup allowed the relaxations due to excess interface volume to be accommodated by the vacuum. In order to obtain the interfacial work of separation ($W_\text{sep}$), the structures were separated by an additional $10\uu{\AA}$ at the designated plane of interest. More information on the tested planes is given in Supplementary Materials Fig.~S1 and Table~S6. Bulk calculations were conducted using a $4\times4\times4$ bcc Fe and a $2\times2\times2$ cementite supercells, each containing 128 atoms, to circumvent the interactions between the solute atom and its periodic boundary image. Fig.~\ref{fig:atomistic_models} shows the most important bulk and slab systems, while Table~\ref{tab:interface_properties} summarizes their structural properties.  
Molecular static calculations followed the same strategy, except that the bulk lattice parameters obtained from full structural relaxations with the respective uMLIPs were used to construct the slab models. For the construction of the semi-coherent interface, we have utilized the \texttt{Interface} module from pymatgen~\cite{ong2013python}, which takes the relaxed bulk structures as an input. Our resulting model contains seven unit cells of cementite fitting eight unit cells of ferrite in the $\left[100\right]_{\text{c}}\parallel \left[\bar110\right]_{\text{f}}$ direction. In the perpendicular interface direction the two phases are fully coherent.

\begin{table}[htb]
    \centering
    \begin{tabular}{c c c c c c}
        \hline
        $N_\text{int}$ & $a$ ($\AA$) & $b$ ($\AA$) & $c$ ($\AA$) & $\gamma_\text{int}$ (J/m$^2$)& $W_\text{sep}$ (J/m$^2$)\\
         \hline
         56 & 4.008 & 4.909 & 46.684 & 0.397 & 4.84 \\
         \hline         
    \end{tabular}
    \caption{Summary of DFT-calculated properties for the coherent interface, including the number of atoms in the slab ($N_\text{int}$), dimensions of the simulation boxes ($a$, $b$ and $c$), corresponding interface energy ($\gamma_\text{int}$), and work of separation ($W_\text{sep}$).}
    \label{tab:interface_properties}
\end{table}

\section{Results}

The results are organized as follows: sections \ref{sec:bulk_energetics} and \ref{sec:seg_coherent} introduce DFT-calculated characteristics of bulk phases and the $\alpha$-Fe/\ce{Fe3C} interface. These are used in the section~\ref{sec:uMLIP_selection} to benchmark uMLIPs and select the best-performing one, which is subsequently used for modeling of the semi-coherent interface in section~\ref{sec:semi-coherent}.

\subsection{Bulk defect energetics}
\label{sec:bulk_energetics}
\begin{figure}[ht!]
    \centering
    \includegraphics[width=0.5\linewidth]{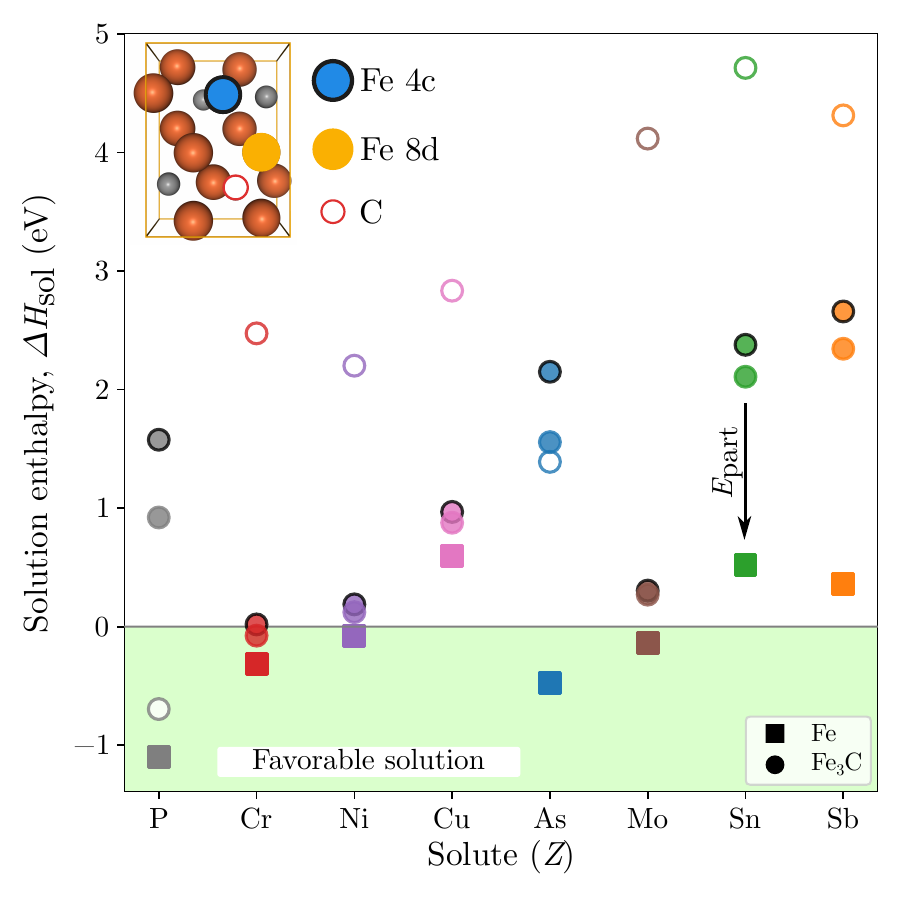}
    \caption{Solution enthalpies, $\Delta H_\text{sol}$, in bcc $\alpha$-Fe and \ce{Fe3C}. The three inequivalent substitutional sites in \ce{Fe3C} are marked in the structural model. The partitioning energy, $E_\text{part}$, quantifies the thermodynamic preference of the solute for the ferrite or cementite phase and is indicated by the black arrow. The green background highlights situations of favorable dissolution.}
    \label{fig:solution_enthalpy_vasp}
\end{figure}

We start by identifying the preferred substitutional sites in bulk as reference states for segregation. Specifically, we investigate the solution enthalpy, $\Delta H_\text{sol}$:
\begin{equation}
    \Delta H_\text{sol} = E_\text{bulk:solute} + \mu_\text{host}  - E_\text{bulk} - \mu_\text{solute}\ ,
\end{equation}
where  $E_\text{bulk:solute}$ and $E_\text{bulk}$ are the total energies of the bulk structure with and without the dissolved species, respectively. The chemical potentials $\mu_\text{host}$ and $\mu_\text{solute}$ correspond to the substituted host atom and solute, respectively (and were set to the total energies per atom of the ground state structure of each species). Consequently, a negative $\Delta H_\text{sol}$ indicates that the solute prefers to dissolve in the host material (i.e., favorable substitution), while a positive $\Delta H_\text{sol}$ indicates that the solute prefers to remain separate (i.e., unfavorable substitution). Unlike ferrite, in which all substitutional sites are crystallographically equivalent, cementite possesses three distinct substitutional sites, as indicated in the structural model in the inset of Fig.~\ref{fig:solution_enthalpy_vasp}.

A markedly different solution behavior is observed between the two bulk phases. While P, As, Cr, Mo, and Ni are soluble in bcc-Fe in descending order of solubility, only P and Cr tend to do so in cementite. Moreover, P prefers to substitute the C atom, whereas Cr tends to replace Fe at the 8d Wyckoff position. These results are in general agreement with existing literature data on the solute impact on the formation energy of the respective phases~\cite{Sakic2022-tt, ande2010first}. In contrast, Cu, Sn, and Sb are the only solutes consistently showing positive $\Delta H_\text{sol}$, in line with their experimentally observed low-temperature immiscibility in $\alpha$-Fe~\cite{kaufman1978coupled, liu2005structure, park2023thermodynamic}. Finally, the difference between the $\Delta H_\text{sol}$ in ferrite and $\min(\Delta H_\text{sol})$ in cementite defines the partitioning energy $E_\text{part}$, which gives the energetically preferred phase for a solute $Z$:
\begin{equation}
    E_\text{part} = \Delta H_\text{sol}(Z,\ce{Fe}) - \min(\Delta H_\text{sol}(Z,\ce{Fe3C}))\ .
\end{equation}
Clearly, all investigated solutes partition into $\alpha$-Fe, indicated by smaller $\Delta H_\text{sol}$, even though in some cases, e.g., Cu, Sn, or Sb, the solution of the solute can be energetically overall unfavorable (i.e., demixing of the solute solution). Although these results agree well with the existing literature~\cite{Sakic2022-tt, ande2010first, sawada2016partitioning}, discrepancies exist between computational $0\,\text{K}$ predictions and experimental measurements for the carbide-forming elements Cr and Mo, which experimentally show partitioning into cementite rather than ferrite~\cite{medouni2020effect}. Previous works have attributed it to the increased magnetic entropy contribution of cementite above its Curie temperature~\cite{sawada2016partitioning}, an effect missing in the $0\uu{K}$ calculations.    

\subsection{Segregation to the coherent interface and effect on cohesion}
\label{sec:seg_coherent}
After identifying the preferred substitutional bulk site, we evaluate the segregation tendencies of the solutes toward the $\alpha$-Fe/\ce{Fe3C} interface with the Bagaryatsky OR. The segregation energy, $E_\text{seg}(X)$, quantifying the driving force for a solute atom to migrate from the bulk to the interface, is calculated using two slightly different expressions, depending on whether the solute substitutes Fe or C atom at the interface:
\begin{equation}
\label{eq:eseg_eq}
    \begin{aligned}
        & E_\text{seg}(X@\ce{Fe}) = E_\text{int}\left[(n-1)\ce{Fe},m\ce{C},X\right] + E_\text{bulk} - E_\text{int} - E_\text{bulk}\left[(u-1)\ce{Fe}, X\right]\ ,\\ 
        & E_\text{seg}(X@\ce{C}) = E_\text{int}\left[n\ce{Fe},(m-1)\ce{C},X\right] + E_\text{bulk} + \mu_{\ce{C}} - E_\text{int} - E_\text{bulk}\left[(u-1)\ce{Fe}, X\right] - \mu_{\ce{Fe}}\ .
    \end{aligned}
\end{equation}
Here, $E_\text{int}\left[(n-1)\ce{Fe},m\ce{C},X\right]$, $E_\text{int}\left[n\ce{Fe},(m-1)\ce{C},X\right]$, and $E_\text{int}$ denote the total energies of the interface structures containing a solute at Fe or C site and the pristine interface, respectively. $E_\text{bulk}\left[(u-1)\ce{Fe}, X\right]$ and $E_\text{bulk}$ represent the total energies of the bulk reference system (ferrite in our case, cf. Fig.~\ref{fig:solution_enthalpy_vasp}) with and without the solute. In cases where the C atom is replaced, the chemical potentials $\mu_{\ce{C}}$ and $\mu_{\ce{fe}}$ are also included to maintain compositional balance. Segregation from the bulk to the interface is thermodynamically favorable when $E_\text{seg} < 0$, whereas positive values indicate solute depletion from the interface.

The mechanical strength of an interface is, based on the Rice-Thomson-Wang model, controlled by the work of separation, $W_\text{sep}$, given by
\begin{equation}
    W_\text{sep} = \frac{E_\text{int}^\text{sep} - E_\text{int}}{A}
\end{equation}
where $E_\text{int}$ and $E_\text{int}^\text{sep}$ are the energy of the intact and separated interface structures, respectively. In this work, we utilize the rigid work of separation, which is obtained without the relaxations of the created surface atoms in the separated interface structure. The effect of the segregated species on the mechanical strength ($\equiv W_\text{sep}$) of the interface is then characterized with the change in cohesion, denoted as $\eta (X)$:
\begin{equation}
    \eta (X) = W_\text{sep}(X) - W_\text{sep}(\emptyset)\ .
\end{equation}
Here, $W_\text{sep}(X)$ and $W_\text{sep}(\emptyset)$ give the $W_\text{sep}$ of the decorated and pristine interface. It follows that a positive (negative) change in cohesion leads to strengthening (weakening) of the interface caused by the segregated solute.

\begin{figure}[ht!]
    \centering
    \includegraphics[width=0.45\textwidth]{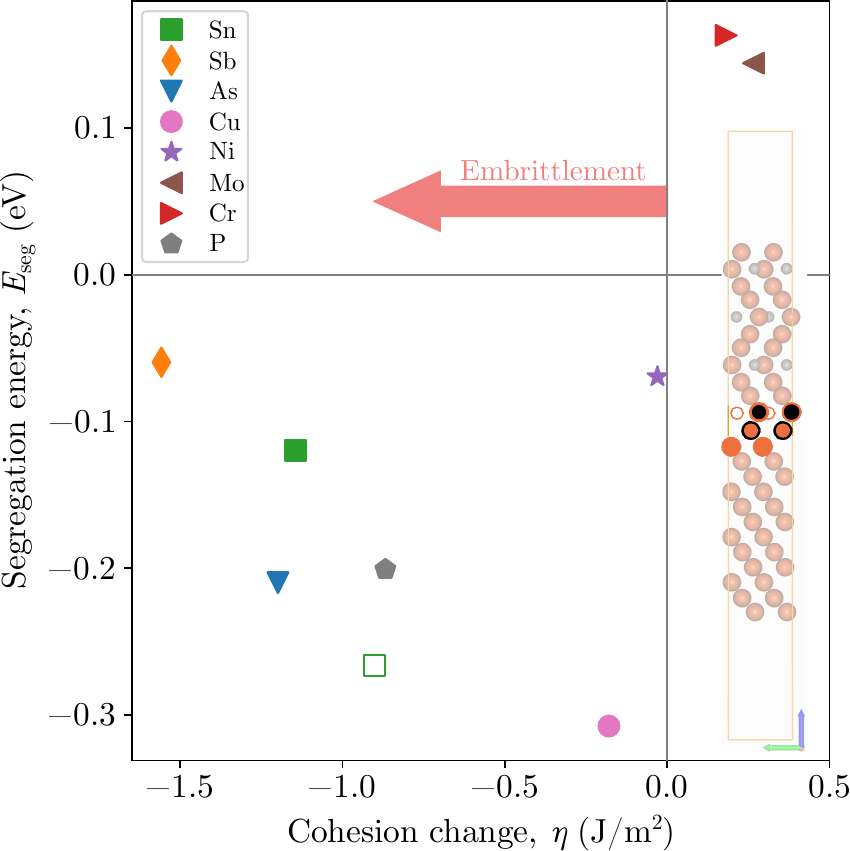}
    \caption{Impact of solute segregation on $\alpha$-Fe/\ce{Fe3C} interface cohesion, $\eta$ (J/m$^2$). For clarity, results are shown only for scenarios with segregation tendency, i.e., $E_\text{seg}(X)\leq 0$ (or minimum $E_\text{seg}(X)$ in case of anti-segregating behavior of Cr and Mo). The explored sites are indicated in the structural model (right). The symbol style for each data point directly corresponds to the highlighted atom position in the displayed model: full symbols (without outline) are Fe sites on the ferrite side of the interface, while the open symbol corresponds to the C site in cementite.}
    \label{fig:dft_eseg_wsep}
\end{figure}

Figure~\ref{fig:dft_eseg_wsep} summarizes the results of segregation behavior and interfacial cohesion. 
All elements, except Sn, segregate to only a single site at the interface, specifically, the Fe atom layer highlighted by the fully orange atoms in the structural model. This indicates that, unlike in Fe GBs, where a distribution of $E_\text{seg}$ values is typically observed~\cite{Sakic2024-ny} even for a single GB, segregation to the $\alpha$-Fe/\ce{Fe3C} interface is confined to only a few energetically favorable site. Consequently, at most a monolayer enrichment is expected at the interface. Sn is the only solute that segregates to two different sites: the same Fe site as all other species, but even more preferably to the C site in the interface region. Both data points are included in the graph. The thermodynamic driving force for segregation follows the sequence $\ce{Cu} > \ce{Sn} > \ce{As}\approx\ce{P} > \ce{Ni} \approx{Sb}$ while Cr and Mo exhibit approximately the same tendency to deplete from the interface. These trends are in good agreement with the segregation behavior observed to Fe GBs, but, importantly, the absolute magnitudes, $|E_\text{seg}|$, are considerably smaller at the $\alpha$-Fe/\ce{Fe3C} interface. For instance, Cu and Sn exhibit segregation energies of $\approx -0.3\,\text{eV}$, compared to $-0.5 < E_\text{seg} < -0.8\,\text{eV}$ for Cu and $\approx -1.3\,\text{eV}$ for Sn at Fe $\Sigma3(1\bar{1}1)[110]$/$\Sigma9(2\bar{2}1)[110]$ GBs~\cite{Sakic2024-ny, mai2022segregation, mai2023phosphorus}. Similar differences exist for the other solutes as well. However, it is noteworthy that the here obtained $E_\text{seg}$ values are comparable to those reported for the special, bulk-like $\Sigma 3(1\bar{1}2)[110]$ GB, which also exhibits significantly lower segregation energies compared to other $\Sigma$ GBs reported in the literature~\cite{Sakic2024-ny, mai2022segregation, mai2023phosphorus}. This is likely due to the dense character of the pearlite interface, similar to the bulk-like GB. Further, this also aligns well with the anti-segregating tendencies of Cr and Mo~\cite{Sakic2024-ny, mai2022segregation, mai2023phosphorus}. 

Looking at the impact on cohesion, the obtained values are not only trend-wise but also magnitude-wise in very good agreement with those reported for Fe GBs. The negligible embrittling effect of Ni is particularly consistent with previous findings, suggesting that the role of Ni could even become cohesion-enhancing in other $\alpha$-\ce{Fe}/\ce{Fe3C} ORs (i.e., indicating an OR-dependent effect on cohesion). The order of cohesion weakening (embrittlement potency) follows $\ce{Sb}>\ce{As}>\ce{Sn}\approx\ce{P}\gg\ce{Cu}>\ce{Ni}$, while Cr and Mo act as cohesion enhancers. However, as calculated in this work, both Cr and Mo are expected to deplete from the interface due to their positive segregation energies ($E_\text{seg} > 0$).
In summary, the segregation tendency and cohesion impact of tramp and trace elements on the coherent pearlite interface is similar to the bulk-like $\Sigma 3(1\bar{1}2)[110]$ GB in $\alpha$-\ce{Fe}.

\subsection{uMLIPs performance and selection}
\label{sec:uMLIP_selection}
\begin{figure}[p!]
    \centering
    \includegraphics[width=0.9\linewidth,height=0.88\textheight,keepaspectratio]{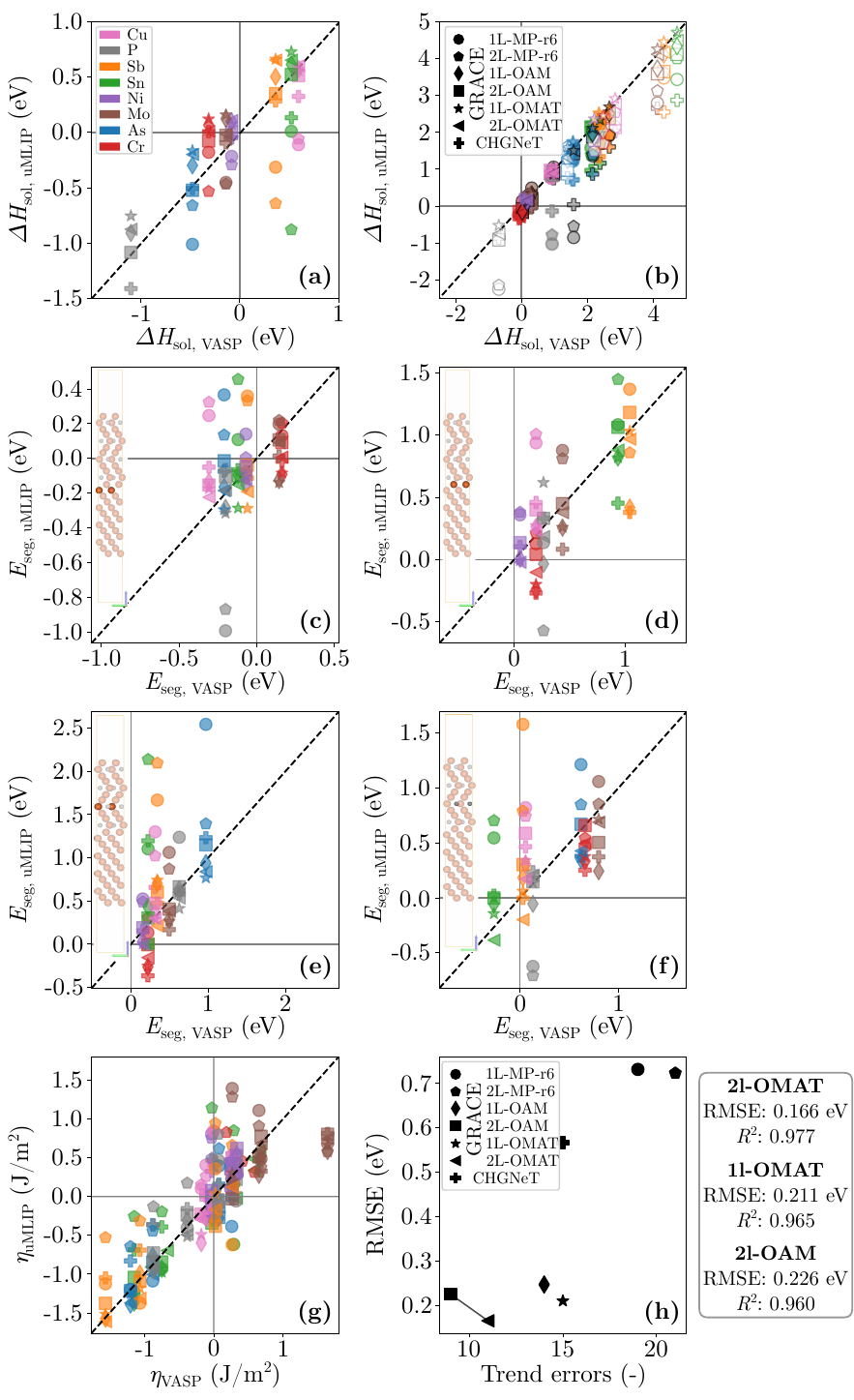}
    \caption{Comparison between VASP and uMLIP predictions. Solute solution enthalpies ($\Delta H_\text{sol}$) in (a) $\alpha$-Fe and (b) \ce{Fe3C}. (c--f) Segregation energies ($E_\text{seg}$) at distinct interfacial sites, as highlighted in the structural models. (g) Cohesion change ($\eta$) corresponding to the $\min(E_\text{seg})$ site, evaluated for different cutting planes. (h) Overall model RMSE against the number of cases where the predicted solute trend (i.e., the sign of the quantity) deviates from DFT (e.g., VASP predicts a favorable solid solution or strengthening effect while uMLIP predicts the opposite). Performance metrics are shown for the best three models.}
    \label{fig:combined_benchmark}
\end{figure}

To identify the most reliable uMLIP model for later modelling of semi-coherent interfaces, all models were benchmarked against DFT reference data for solution enthalpies, segregation energies, and interfacial cohesion effects (Fig.~\ref{fig:combined_benchmark}). Model performance was assessed using complementary metrics, including the root-mean-squared error (RMSE) and the coefficient of determination ($R^2$); their values for the three best-performing models are given in Fig.~\ref{fig:combined_benchmark}h. However, these metrics alone do not reveal whether a model correctly captures the qualitative trends observed with DFT, e.g., whether a solute prefers to dissolve in a given bulk phase or not, or if it tends to segregate/deplete to/from a specific site at the interface. To address this, Fig.~\ref{fig:combined_benchmark}h presents a Pareto-style comparison of each model's RMSE against the number of cases in which the predicted trend opposes that from DFT.
% \newpage
Among the uMLIPs, the GRACE-MP-r6 models exhibit in many cases large energetic deviations and hence also in many cases predict the wrong physical trend (Fig.~\ref{fig:combined_benchmark}h). For example, the preferable solution of Sn and Sb in $\alpha$-Fe (Fig.~\ref{fig:combined_benchmark}a), P in \ce{Fe3C} (Fig.~\ref{fig:combined_benchmark}b), or anti-segregation of Sn, Sb, As, Cu (Fig.~\ref{fig:combined_benchmark}c), to name a few. The CHGNet model, although the only model capable of predicting the magnetic moments, takes an intermediate position with slightly increased accuracy compared to the GRACE-MP-r6 models, and also improved trend alignment. Interestingly, the CHGNet predictions are particularly wrong whenever the solute is placed on a C position (Fig.~\ref{fig:combined_benchmark}b) or in the vicinity of the C atom (Fig.~\ref{fig:combined_benchmark}d--f).
At the opposite end of the performance spectrum lie the GRACE-OMAT and GRACE-OAM models, which achieve the lowest overall errors and most closely reproduce the DFT results. Surprisingly, the GRACE-2L-OMAT and GRACE-1L-OMAT models yield even lower RMSE values than the OAM variants, despite the fact that the training set of the latter is larger. This performance gap appears to stem primarily from the larger energetic deviations associated with C-substituted $\Delta H_\text{sol}$ in cementite (Fig.~\ref{fig:combined_benchmark}b) and C-substituted $E_\text{seg}$ at the interface (Fig.~\ref{fig:combined_benchmark}f). Conversely, for $\Delta H_\text{sol}$ in ferrite (Fig.~\ref{fig:combined_benchmark}a), the GRACE-2L-OAM model outperforms both OMAT variants and essentially matches the DFT results. Moreover, the GRACE-2L-OAM matches the DFT determined trends (qualitative agreement between DFT and uMLIP) the best, followed by the GRACE-2L-OMAT model, which in turn puts both models closest to the origin of the Pareto-style chart in Fig.~\ref{fig:combined_benchmark}h. Our tests also reveal that Cr as a solute produces the largest fraction of trend inconsistency for both models.
In summary, the GRACE-2L-OMAT and GRACE-2L-OAM models perform comparably well, and both represent options for addressing the underlying research question related to segregation in pearlite. Given that we prioritize consistency with DFT-based trends---and therefore the physical interpretability of the results---over a marginal RMSE difference of $0.06\,\text{eV}$, the GRACE-2L-OAM potential was selected for further investigations. Finally, it is worth noting that the observed $E_\text{seg}$ deviations between the two best GRACE models and DFT remain within the range of $E_\text{seg}$ values reported in the literature from independent DFT studies.

The uMLIPs also allow for easily evaluating models with sizes beyond what is routinely feasible with DFT. Therefore, we have probed segregation behaviour using the GRACE-2L-OAM uMLIP to a laterally extended model ($8\times$ in the $[100]_\text{c} \| [\bar110]_\text{f}$ direction)  containing overall 672 atoms. The results are reported in Supplementary Material (Fig.~S2) and clearly show that there is no important difference between the ``small'' and ``big'' models. Both agree qualitatively, and, importantly, for segregating or only slightly anti-segregating scenarios, also quantitatively, with only differences obtained in the strongly anti-segregating cases.

\subsection{Segregation to semi-coherent interface}
\label{sec:semi-coherent}
\begin{figure}[ht!]
    \centering
    \includegraphics[width=0.3\textheight]{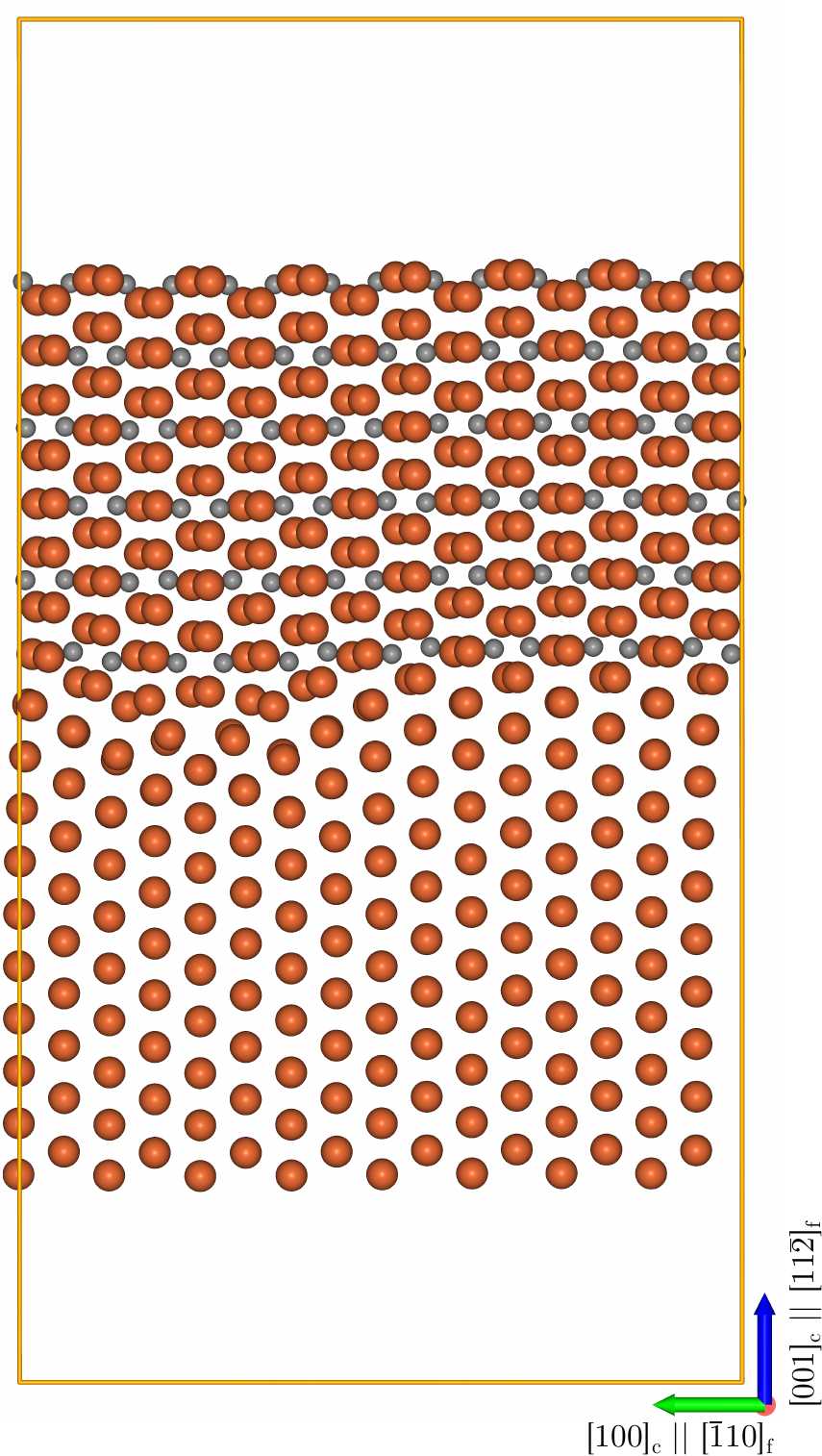}
    \caption{Relaxed semi-coherent $\alpha$-\ce{Fe}/\ce{Fe3C} interface following the Bagaryatsky OR.}
    \label{fig:semicoherent_interface_dxa}
\end{figure}

The semi-coherent interface following the Bagaryatsky OR after structural relaxation using the GRACE-2L-OAM uMLIP is shown in Fig.~\ref{fig:semicoherent_interface_dxa}. 

As can be seen from Fig.~\ref{fig:semicoherent_interface_dxa}, the misfit dislocation corresponds to an edge dislocation along the $[111]_f$ direction with Burgers vector $\vec{b}=a[\bar110]_f$, corresponding to a magnitude of $|\vec{b}|=4.01\,\AA$. The dislocation is a direct consequence of the employed model (7 cementite unit cells fit to 8 ferrite unit cells according to the Bagaryatsky OR). The resulting local relaxations near the dislocation core are visualized in Supplementary Material Fig.~S3. The core of the misfit dislocation at the interface produces atomic arrangements extending into the ferrite. High-resolution experimental studies of pearlitic steels revealed the existence of interface steps associated with the development of lamellar curvature without any change in the habit plane of the interface~\cite{Zhou1992-jl}. We propose that both cases are related, as both serve to locally relax the lattice mismatch between ferrite and cementite.

\begin{figure}
    \centering
    \includegraphics[width=1\textwidth]{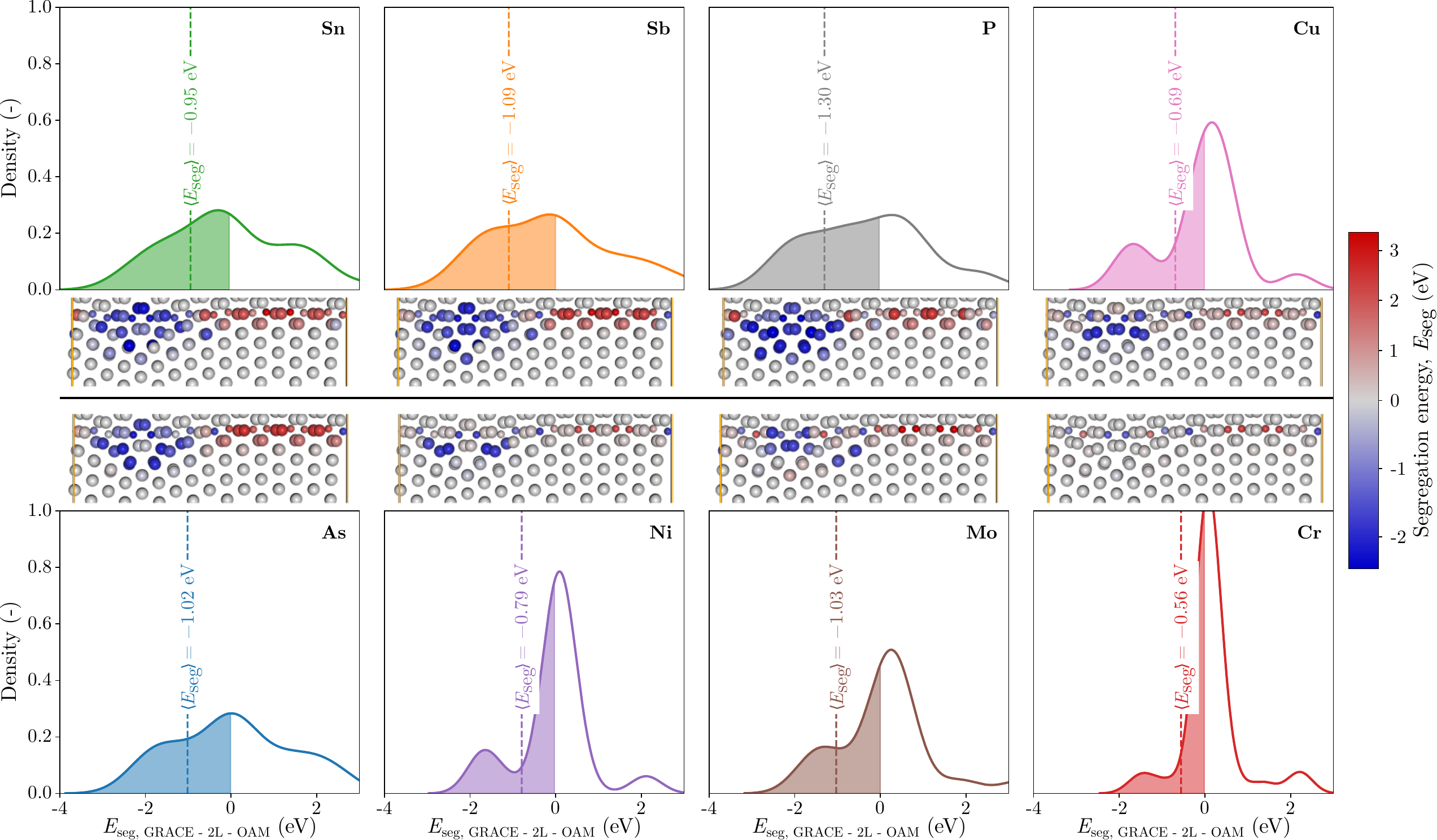}
    \caption{Upper and lower panels display normalized histograms of the segregation energies, $E_\text{seg}$, to the semi-coherent $\alpha$-\ce{Fe}/\ce{Fe3C} interface. Dashed lines indicate the mean negative segregation energy, $\left<E_\text{seg}^{-}\right>$. The middle panels display the atomic sites within the semi-coherent interface, colored according to their site-specific $E_\text{seg}$.}
    \label{fig:semicoherent_eseg_distr}
\end{figure}

The segregation energy spectra of the individual solutes, along with their site-resolved $E_\text{seg}$ in the semi-coherent interface, are shown in Fig.~\ref{fig:semicoherent_eseg_distr}. 
Evidently, the semi-coherent interface acts as a pronounced segregation sink for solute atoms, with $E_\text{seg}$ values reaching below $\approx -1.5\uu{eV}$ for all investigated elements. These energies are significantly lower than those in the coherent interface, where Cu exhibited the strongest segregation tendency with $E_\text{seg} \approx -0.3\uu{eV}$. The site-resolved segregation maps in Fig.~\ref{fig:semicoherent_eseg_distr} further reveal that the deepest traps for all solutes are located in the vicinity of the dislocation core. However, the spatial distribution of $E_\text{seg}$ around the dislocation varies markedly between the solute species. For instance, Sn, Sb, P, and As exhibit pronounced segregation in a spatially extended area around the dislocation, with multiple favorable binding sites. The order of segregation propensity, as determined from the mean negative segregation energy, $\left<E_\text{seg}^{-}\right>$, follows $\ce{P}>\ce{Sb}\gtrapprox \ce{As}\gtrapprox \ce{Sn}$.
In contrast, solutes such as Mo, Ni, and Cu display potent trapping only in the vicinity of the dislocation core, with their mean negative $\left<E_\text{seg}^{-}\right>$ values increasing in the same sequence. Cr shows a distinct behavior, being attracted primarily to the C atoms adjacent to the dislocation core.
It is worth noting that these trends roughly follow the atomic sizes: P with $98\uu{pm}$ being the smallest of these atoms, followed by As, Sb, and Sn. For the latter two, Sb and Sn, the segregation is likely significantly enhanced by their insolubility in both ferrite and cementite phases (cf. Fig.~\ref{fig:solution_enthalpy_vasp}). The stronger Mo segregation to the dislocation core can also be related to its large radius ($190\uu{pm}$ as compared to $156\uu{pm}$ for Fe)

Beyond these element-specific trends and patterns, three characteristic features can be identified in the overall $E_\text{seg}$ spectra. The first, a peak in the negative regime, corresponds to the deep segregation traps surrounding the dislocation center. The second, smaller hump in the positive energy range (comparable in magnitude only for Cr), arises from sites located midway between the dislocation and its periodic image. The third and most populated peak, centered around $0\uu{eV}$, represents atomic sites situated between adjacent dislocations but displaced by one to two atomic layers in the perpendicular direction from the interface plane.

\begin{figure}[htb]
    \centering
    \includegraphics[width=0.45\textwidth]{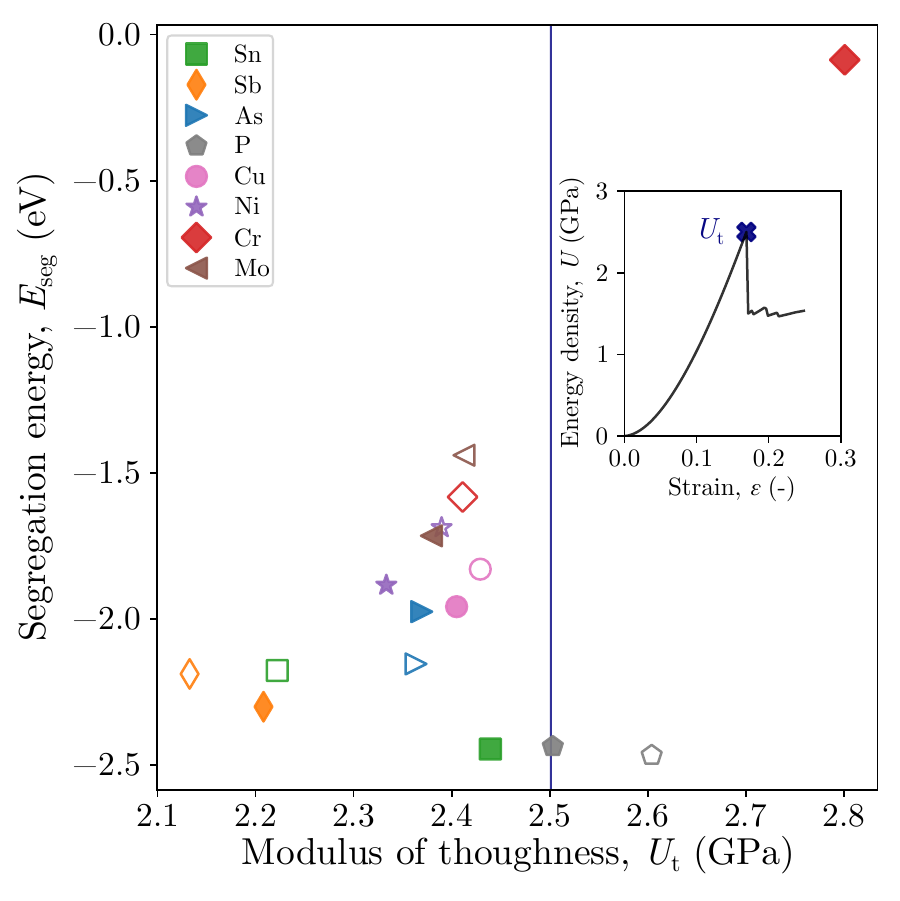}
    \caption{Impact of solute segregation on the semi-coherent interfacial modulus of toughness, $U_\text{t}$, describing the maximum absorbed energy up to the ultimate tensile strength. The latter is marked in the inset plot which shows the energy density versus strain curve for the undecorated semi-coherent interface. Solid and open markers indicate substitution of the $\min(E_\text{seg})$–Fe and $\min(E_\text{seg})$–C sites, respectively, and the vertical blue line indicates $U_\text{t}$ of the undecorated interface.}
    \label{fig:UT_semicoh}
\end{figure}

To quantify the mechanical impact of solute segregation on the semi-coherent interface, tensile tests perpendicular to the interface plane were performed. Two solute positions were considered: each representing the respective $\min(E_\text{seg})$ of Fe and C sites. Fig.~\ref{fig:UT_semicoh} shows the resulting changes in the modulus of toughness, $U_\text{t}$, defined here as the area under the stress-strain curve up to the ultimate tensile strength. It should be noted that $U_\text{t}$, as used in this work, does not represent the fracture toughness, which instead describes the critical stress intensity factor required to propagate an existing crack.
As can be seen in Fig.~\ref{fig:UT_semicoh}, most solutes decrease $U_\text{t}$, indicating a reduction in the maximum strain energy that the interface can accommodate under tension. We note that the $U_\text{t}$ cannot be directly compared to the work of separation due to the role of atomic relaxations. 

The calculated $U_\text{t}$ point to an overall embrittling effect for the majority of investigated solute configurations. Two notable exceptions are observed. First, Cr exhibits a site-dependent influence, i.e. while segregation to the Fe-site increases $U_\text{t}$ relative to the undecorated interface, $U_\text{t}$ is reduced when segregating to the C site. Nonetheless, since segregation to the C-site is significantly stronger, Cr is expected to embrittle the semi-coherent interface. Second, P does slightly increase $U_\text{t}$ for the preferred segregation site (C) or does not alter $U_\text{t}$ for segregation to the preferred Fe site, and thereby is the only non-embrittling element among those here considered. 

In contrast to the coherent interface, where Mo enhances cohesion and Ni has little influence, both Mo and Ni promote embrittlement when segregated to regions near the misfit dislocation. A similar degree of embrittlement is predicted for Cu and As. The most pronounced reduction in $U_\text{t}$ is found for Sn and, in particular, Sb, which exhibit both strong segregation tendencies toward the dislocation core and the largest detrimental effect on the strain energy that can be absorbed by the interface.

\section{Discussion}
\subsection{Segregation behavior: comparison with experiment}
The calculated $E_\text{seg}$ values close to zero at the coherent interface imply that the solute excess is rather minor and hence may be hardly measurable. In pure pearlitic steels, the segregation enrichment is expected to be low under thermodynamic equilibrium. However, it remains uncertain whether such equilibrium can be fully established or is hindered by the specific processing parameters of steels.

Examining the segregation energetics obtained for the coherent $\alpha$-Fe/\ce{Fe3C} interface (Fig.~\ref{fig:dft_eseg_wsep}), three distinct groups with different $E_\text{seg}$ ranges can be identified. The first group---exhibiting positive $E_\text{seg}$ values---includes Cr and Mo, which act as depleting solutes. The second group, consisting of Sb, Ni, and Sn(@Fe), shows weak segregation tendencies with $E_\text{seg}$ values around $\geq -0.12\uu{eV}$, and thus these solutes are not expected to produce a measurable interfacial excess in experiments. The final group, with moderately negative segregation energies ($\approx -0.3 < E_{seg} \lessapprox -0.2$ eV), includes As, P, Sn(@C), and Cu, showing increasing segregation affinity in this order and being expected to yield measurable enrichment at the interface. Indeed, P segregation has been observed experimentally by atom probe tomography (APT) in low-alloy steels~\cite{medouni2020effect}. Therefore, we propose that also the other solutes in this group could also be experimentally detected at the interface.

In non-pure pearlitic steels, however, segregation to the $\alpha$-Fe/\ce{Fe3C} interface competes with segregation to Fe grain boundaries (GBs). As discussed in Sec.~\ref{sec:seg_coherent}, the segregation energies to Fe GBs are more negative, implying that solutes preferentially enrich GBs. The reality is further complicated by possible co-segregation, as well as repulsive or attractive solute–solute interactions~\cite{Sakic2024-ny}, which in the latter case can promote precipitate formation at the interface. STEM analysis in Ref.~\cite{duan2024effect} revealed the formation of Cu precipitates not only along Fe GBs but also prominently at pearlitic interfaces. It can be assumed that these precipitates formed following prior Cu segregation to the interface. The same study also reported Ni and Sn enrichment in the vicinity of these precipitates; our simulations likewise identify these elements as segregating to the $\alpha$-Fe/\ce{Fe3C} interface. Similarly, Sb and Ni were found at Cu-rich precipitates in a low-alloy steel~\cite{li2023role}. These reports suggest that attractive interfacial interactions among Cu, Sb, Sn, and Ni may lead to local accumulation, thereby reducing solute excess in other regions of the interface, potentially to levels below the detection limits of current experimental techniques.

Shifting our attention to the semi-coherent interface, we find a drastically different picture. All solute elements, without exception, segregate strongly into the vicinity of the misfit dislocation. Although the number of energetically favorable sites varies among solutes, minimum segregation energies of $E_\text{seg} < -1.5\uu{eV}$ are observed for all cases. This exceeds not only the segregation strengths found at the coherent interface but also those reported in the literature for Fe GBs~\cite{Sakic2024-ny, mai2022segregation}.

Our findings indicate that the here studied solute elements strongly segregate to the dislocation core regions at the semi-coherent interface.
These predictions are in line with the conclusions drawn in Ref.~\cite{Yu2019-cu}, where H was identified to segregate preferentially to defect-rich regions of the $\alpha$-Fe/\ce{Fe3C} interface. Similarly, Niu \etal~\cite{Niu2024-vz} observed that the plastic deformation behavior during micropillar compression changed from shearing along $\alpha$-Fe/\ce{Fe3C} lamellae in H-uncharged samples to slip within the ferrite in H-charged samples. The authors proposed the H-enhanced dislocation mobility in ferrite. However, this transition may also be explained by increased H segregation to the misfit dislocation core region and subsequent dislocation pinning, which may shift plastic deformation into the ferrite interior. While this interpretation remains speculative, it aligns with the conclusions drawn in Ref.~\cite{Yu2019-cu} and highlights an avenue for future investigations enabled by uMLIPs.

\subsection{Selection of the appropriate uMLIP}

As demonstrated in the present work, uMLIPs are capable of accurately treating multi-component systems within structurally complex atomic environments, states far from what they have been trained on. Nevertheless, their performance varies, underscoring the necessity of rigorous benchmarking. In our study, two potentials stood out: GRACE-2L-OMAT and GRACE-2L-OAM. The choice of GRACE-2L-OAM for modeling the semi-coherent interface was based on its slightly better trend alignment with DFT results---i.e., consistent signs for corresponding physical quantities ($+\Delta H_\text{sol, VASP} \rightarrow +\Delta H_\text{sol, uMLIP}$, $-\Delta H_\text{sol, VASP} \rightarrow -\Delta H_\text{sol, uMLIP}$, etc.)---across bulk solution energetics, segregation energies, and work of separation at the coherent interface. Although the GRACE-2L-OMAT potential exhibited a marginally smaller root-mean-square error ($\text{RMSE} = 0.166\,\text{eV}$ compared to $0.226\,\text{eV}$) obtained for the GRACE-2L-OAM (see Fig.~\ref{fig:combined_benchmark}h), the trend consistency (i.e., qualitity=segregating or anti-segregating, rather than quantity=how strongly (anti-)segregating) favored the latter one. 

This comparison illustrates two important points. Firstly, the community needs a clear consensus on how to select the most appropriate uMLIP for a given application. To address this issue, the Matbench Discovery Scoreboard~\cite{Riebesell2025-ar} introduced the Combined Performance Score ($\text{CPA}$), which weights discovery performance ($F_1$), geometry optimization accuracy (RMSD), and thermal conductivity prediction error ($\kappa\text{SRME}$ ). However, such a composite metric may not be universally applicable, as in our case, where thermal conductivity is irrelevant to the studied problem. Secondly, not all uMLIPs suitable for specific materials science problems, such as GRACE-2L-OMAT in the present work,are necessarily included in widely known databases. This underscores that the best-performing potential for a particular system and/or problem may not yet be listed in such repositories. For completeness, we have also computed the full segregation spectra for the semi-coherent interface using the GRACE-2L-OMAT potential and compared them with the GRACE-2L-OAM results (Fig.~\ref{fig:semicoherent_eseg_distr}). The obtained $E_\text{seg}$ distributions are in quantitative agreement, and hence are only shown in Supplementary Material Fig.~S4.

\subsection{Impact on interface strength}

Returning to the discussion of solute segregation, we find that the impact on the mechanical strength of the $\alpha$-Fe/\ce{Fe3C} interface is quite similar to what has been reported for $\alpha$-Fe GBs in Ref~\cite{Sakic2024-ny}. The coherent interface becomes strongly embrittled by Sb, Sn, As, and P, and only mildly by Cu and negligibly by Ni. The magnitudes of cohesion loss induced by the individual solutes are comparable to those reported in Ref.~\cite{Sakic2024-ny} for the $\Sigma3(1\bar{1}2)[110]$ GB, which is characterized by low excess volume and often described as a stacking-fault or twin-like representative~\cite{mai2022segregation}. 

An embrittling tendency is also found in the semi-coherent interface for all solutes except P. Particularly interesting in this context are Mo and Cr, which in the coherent interface were found to increase cohesion. However, this is unlikely to have any practical consequence, as Cr and Mo tend to deplete from the coherent interface. While enhancing cohesion, the impact of P is only marginal. It is noted that in this work only tensile tests perpendicular to the interface plane were performed. Hence, a change of the deformation mode, e.g. into in-plane shearing, may change the material response due to blocking of the dislocation movement. Overall, also in the semi-coherent interface the influence of Sn, and especially Sb, show the most detrimental effect, as both solutes strongly segregate and reduce the strain energy that the system can absorb. It is therefore reasonable to assume that, with the increasing use of recycled steels, these solutes will not only segregate to and reduce the cohesion of $\alpha$-Fe GBs but also that of pearlitic $\alpha$-Fe/\ce{Fe3C} interfaces. 

As discussed above, Cu precipitates, often enriched with Sn, Sb, and Ni, are frequently observed at the pearlitic interfaces~\cite{li2023role, duan2024effect}. This underscores the importance of investigating the influence of such precipitates on interfacial mechanical properties and cohesion in order to obtain a deeper understanding of the full picture. Notably, the enhanced segregation affinity of Cu, Sn, Sb, and Ni to the misfit dislocation region makes this area a preferential nucleation site for Cu-rich precipitates. This, in turn, raises an important question regarding the potential interactions between these precipitates and the motion of misfit dislocations during straining, which could have significant implications for the deformation behavior of pearlitic steels. 

\subsection{Note on temperature effects and dilute limit}

Lastly, we note that all results presented in this study were obtained from $0\uu{K}$ DFT and molecular statics calculations. Finite-temperature simulations~\cite{tanaka2020prediction} including phonon and magnetic contributions in bcc Fe have shown that their combined effect on the total free energy remains below $20\,\text{meV/atom}$ in the range of $0\text{--}800\,\text{K}$. However, as discussed in Sec.~\ref{sec:bulk_energetics}, both Cr and Mo partition into ferrite at $0\uu{K}$ rather than into cementite, contrary to what is typically assumed under practical heat-treatment conditions. This discrepancy does not reflect a deficiency of the here shown DFT or uMLIP calculations but rather arises because of the increased magnetic contribution in the partitioning process, as has been shown for the case of Cr. Since we define the ferrite bulk as the reference state for solute segregation, corresponding to the calculated $\Delta H_\text{sol}$ at $0\,\text{K}$, the segregation trends obtained here for Cr and Mo should be interpreted with this reference in mind. At finite temperatures, the bulk reference energy $E_\text{bulk}$ in Eq.~\ref{eq:eseg_eq} would change, which could alter the quantitative segregation values. A detailed treatment of these effects, however, goes beyond the scope of the present work and is left for future investigations, possibly enabled by using (u)MLIPs.

Finally, we note that the results presented here were obtained in the dilute limit, such that solute–solute interactions are not explicitly considered. While the calculated segregation energetics provide a first-order thermodynamic driving force for segregation, especially at low solute concentrations, deviations may arise at higher local concentrations due to solute–solute interactions. In particular, in the vicinity of the misfit dislocation core, these interactions may modify the segregation energetics of subsequent solutes. A quantitative treatment of non-dilute effects, however, would require explicit consideration of multi-solute configurations (e.g., as in Ref.~\cite{Svoboda2020-tr} or via employing Calphad formalism proposed in Ref.~\cite{Spitaler2025-vb}), which is beyond the scope of the present work and is left for future investigations.

\section{Conclusions}

In conclusion, we have demonstrated that current state-of-the-art uMLIPs are sufficiently robust to be employed not only as triaging tools for efficient allocation of computational resources in high-throughput DFT workflows but also as predictive models for addressing specific, physically relevant questions in materials science. 

Using a combined DFT/uMLIP-guided framework, we analyzed the segregation behavior of As, Cu, Cr, Mo, Ni, P, Sb, and Sn at both coherent and semi-coherent $\alpha$-Fe/\ce{Fe3C} interfaces and evaluated their influence on interfacial cohesion. The coherent interface exhibits weak to moderate segregation tendencies, with a minimum $E_\text{seg}\approx -0.3\,\text{eV}$ observed for Cu. In contrast, the semi-coherent interface containing a misfit dislocation acts as a deep trap for all investigated solutes, though the number of attractive sites (i.e., the width of the $E_\text{seg}$ spectrum) is solute-dependent. 

These findings reveal that the structurally complex semi-coherent interface represents a strong segregation sink, raising new important questions about whether (i) the sites in the vicinity of the misfit dislocation act as preferential nucleation centers for precipitates. Further, (ii) how solute segregation interacts with the motion of the misfit dislocations, and (iii) how they may influence the deformation mode of pearlitic interfaces. 

\backmatter

\bmhead{Author contributions}
A.R.S. wrote the main manuscript, performed the calculations and prepared the figures. D.H. wrote parts of the manuscript, supervised and provided the computational resources. All authors reviewed the manuscript.

\bmhead{Acknowledgements}
This work was partly performed using supercomputer resources provided by the Vienna Scientific Cluster (VSC). 

\bmhead{Funding}
The financial support by the Austrian Federal Ministry of Economy, Energy and Tourism, the National Foundation for Research, Technology and Development and the Christian Doppler Research Association is gratefully acknowledged. Part of this research was funded by the Austrian Science Fund (FWF) [P 34179-N].

\bmhead{Competing interest}
The authors declare that they have no known competing financial interests or personal relationships that could have appeared to influence the work reported in this paper.

\bmhead{Data availability}
Data is available from the corresponding author upon request.

\bmhead{Supplementary information}

If your article has accompanying supplementary file/s please state so here. 

Authors reporting data from electrophoretic gels and blots should supply the full unprocessed scans for key as part of their Supplementary information. This may be requested by the editorial team/s if it is missing.

Please refer to Journal-level guidance for any specific requirements.

% \section*{Acknowledgments}

% \section*{Funding declaration}

% \newpage
% \appendix

% \newpage
 % \bibliographystyle{elsarticle-num} 
 \bibliography{refs}

\end{document}